\documentstyle[prl,aps,epsf,multicol]{revtex}
\begin{document}

\title{Anomalous magneto-oscillations in two-dimensional systems}

\draft

\author{R. Winkler}
\address{Institut f\"ur Technische Physik III, Universit\"at
Erlangen-N\"urnberg, Staudtstr. 7, D-91058 Erlangen, Germany}

\author{S. J. Papadakis, E. P. De Poortere, and M. Shayegan\cite{lmu}}
\address{Department of Electrical Engineering, Princeton University,
Princeton, New Jersey 08544}

\date{October 1, 1999}
\maketitle
\begin{abstract}
  The frequencies of Shubnikov-de Haas oscillations have long been
  used to measure the unequal population of spin-split
  two-dimensional subbands in inversion asymmetric systems.  We
  report self-consistent numerical calculations and experimental
  results which indicate that these oscillations are not simply
  related to the zero-magnetic-field spin-subband densities.
\end{abstract}

\begin{multicols}{2}
  \par\global\columnwidth20.5pc
  \global\hsize\columnwidth\global\linewidth\columnwidth
  \global\displaywidth\columnwidth

Spin degeneracy of the electron states in a solid is the combined
effect of inversion symmetry in space and time. Both symmetry
operations change the wave vector ${\bf k}$ into $-{\bf k}$, but
time inversion also flips the spin so that combining both we have a
two-fold degeneracy of the single particle energies, ${\cal
E}_{+}({\bf k}) = {\cal E}_{-}({\bf k})$ (Ref.\ \cite{kit63}). When
the potential through which the carriers move is inversion
asymmetric however, the spin-orbit interaction removes the spin
degeneracy even in the absence of an external magnetic field $B$.
This $B=0$ spin splitting is the subject of considerable interest
because it concerns details of energy band structure that are
important in both fundamental research and electronic device
applications
(\cite{sto83,wol89,luo90,nit97,lu98,pap99,bych84,eke85,ros89,dre92,%
jus92,and94}
and references therein).

The spin splitting of the single particle energies yields two spin
subbands with different populations $N_\pm$. The frequencies
$f_\pm^{\rm\,SdH}$ of longitudinal magnetoresistance oscillations in
small magnetic fields perpendicular to the plane of the system,
known as Shubnikov-de Haas (SdH) oscillations, have often been used
\cite{sto83,wol89,luo90,nit97,lu98,pap99} to measure the $B = 0$
spin-subband densities $N_\pm$ following
\begin{equation}
\label{onsag}
N_\pm = \frac{e}{h} f_\pm^{\rm\,SdH} .
\end{equation}
Here $e$ is the electron charge and $h$ is Planck's constant. Eq.\
(\ref{onsag}) is based on a well-known semiclassical argument due to
Onsager \cite {ons52} which relates the cyclotron motion at $B>0$
with extremal cross sections of the Fermi surface at $B=0$. In this
paper, we test both experimentally and theoretically the validity of
this procedure. We obtain good agreement between experimental and
calculated SdH oscillations. On the other hand the calculated $B=0$
spin splitting differs substantially from the predictions of Eq.\
(\ref{onsag}). We will show that this difference reflects the
inapplicability of conventional Bohr-Sommerfeld quantization for
systems with spin-orbit interaction.

The subject of our investigation are two-dimensional (2D) hole
systems in modulation-doped GaAs quantum wells (QW's).  We use GaAs
because high quality samples can be grown which allow the
observation of many SdH oscillations, and because the band-structure
parameters are well known \cite{may91,lan82} so that accurate
numerical calculations can be performed. The crystal structure of
GaAs is zinc blende, which is inversion asymmetric. Furthermore, a
QW structure can be made asymmetric if an electric field $E_{\perp}$
is applied perpendicular to the plane of the well.  Therefore, at a
given 2D hole density, the $B = 0$ spin splitting in these systems
has a fixed part due to the bulk inversion asymmetry (BIA), and a
tunable part due to the structure inversion asymmetry (SIA).

\begin{figure}
\centerline{\epsfxsize=0.70\columnwidth\leavevmode
 \epsffile{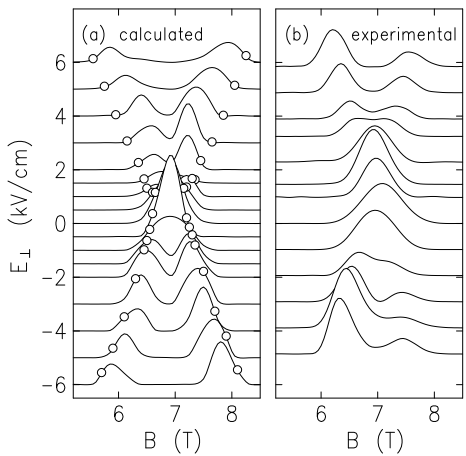}}\vspace{3mm}
\caption[]{\label{pic:w190:surf} Fourier spectra of the (a)
calculated and (b) measured SdH oscillations versus magnetic field
$B$ for different values of electric field $E_\perp$ for a 200 {\AA}
wide GaAs-Al$_{0.3}$Ga$_{0.7}$As QW with growth direction [113] and
2D hole density $N = 3.3 \cdot 10^{11}$~cm$^{-2}$. The open circles
show the expected Fourier transform peak positions $(h/e)N_\pm$
according to the calculated spin splitting $N_\pm$ at $B=0$.}
\end{figure}

Figure \ref{pic:w190:surf} highlights the main findings of our
paper. It shows the Fourier spectra of the calculated [Fig.\ 
\ref{pic:w190:surf}(a)] and measured [Fig.\ \ref{pic:w190:surf}(b)]
SdH oscillations as well as the expected peak positions $(h/e)N_\pm$
according to the calculated spin split densities $N_\pm$ at $B=0$
[open circles in Fig.\ \ref{pic:w190:surf}(a)] for a 2D system with
constant hole density $N = N_{+} + N_{-} = 3.3 \cdot 10^{11}$
cm$^{-2}$ but with varying $E_{\perp}$. Even around $E_\perp = 0$,
when we have only BIA but no SIA, the open circles indicate a
significant spin splitting $\Delta N = N_+ - N_-$. However, the
Fourier spectra in Figs.\ \ref{pic:w190:surf}(a) and (b), while in
good agreement with each other \cite{totdens,asym}, deviate
substantially from the zero-$B$ spin splitting: for nearly all
values of $E_\perp$ the splitting $(h/e)\Delta N$ is significantly
larger than $\Delta f = f_+^{\rm\,SdH} - f_-^{\rm\,SdH}$. In
particular, near $E_\perp = 0$ only one SdH frequency is visible in
both the measured and calculated spectra, whereas we would expect to
obtain two frequencies \cite{beat}.  In the following we will show
how one can understand these results. We will briefly describe some
details of our calculations and experiments and then discuss the
physical origin of when and why Eq.\ (\ref{onsag}) fails.

Our calculations are based on the methods discussed in Refs.\
\cite{win93,rw931,win96}. A multiband Hamiltonian \cite{trr79}
containing the bands $\Gamma_6^c$, $\Gamma_8^v$ and $\Gamma_7^v$ is
used to calculate hole states in the QW. It fully takes into account
the spin splitting due to BIA and SIA. The Poisson equation is
solved self-consistently in order to obtain the Hartree potential.
We obtain two spin-split branches of the energy dispersion ${\cal
E}_\pm ({\bf k}_\|)$ as a function of in-plane wave vector ${\bf
k}_\|$. However, we do not call these branches spin-up or spin-down
because the eigenstates are not spin polarized, i.e., they contain
equal contributions of up and down spinor components. (This reflects
the fact that for $B=0$ the system has a vanishing magnetic moment.)
From ${\cal E}_\pm ({\bf k}_\|)$ we obtain the population $N_\pm$ of
these branches \cite{rw931}.

For the calculation of SdH oscillations we use the very same
Hamiltonian \cite{trr79} discussed above so that the results for
$B=0$ and $B>0$ are directly comparable. We introduce the magnetic
field by replacing the in-plane wave-vector components with Landau
raising and lowering operators in the usual way \cite{win96,trr79}.
From the Landau fan chart, using a Gaussian broadening, we obtain
the oscillatory density of states at the Fermi energy which is
directly related to the electrical conductivity \cite{and74}. In
order to match the experimental situation the Fourier spectra in
Fig.\ \ref{pic:w190:surf}(a) were calculated for $B$ between 0.20
and 0.85~T ($B^{-1}$ between 1.17 and 5.0~T$^{-1}$). We note that
the positions of the peaks in the Fourier spectra in Fig.\ 
\ref{pic:w190:surf}(a) depend only on the Landau fan chart as
determined by the multiband Hamiltonian \cite{gauss}. A single peak
in the Fourier spectrum corresponds to the situation that at the
Fermi energy the spacing between Zeeman-split Landau levels is a
fraction $\alpha$ of the spacing between Landau levels with adjacent
Landau quantum numbers $n$ and $n+1$, with a constant $\alpha$
independent of~$B$.

For measurements, we use Si modulation doped GaAs QW's grown by
molecular beam epitaxy on the (113)A surface of an undoped GaAs
substrate. The well width of the sample in Fig.\ \ref{pic:w190:surf}
is 200 \AA. Photolithography is used to pattern Hall bars for
resistivity measurements.  The samples have metal front and back
gates that control both the 2D hole density and $E_{\perp}$.
Measurements are done at a temperature of 25~mK. In order to vary
$E_{\perp}$ while maintaining constant density we first set the
front gate ($V_{\rm fg}$) and back gate ($V_{\rm bg}$) biases and
measure the resistivities as a function of $B$. The total 2D hole
density $N$ is deduced from the Hall coefficient. Then, at small
$B$, $V_{\rm fg}$ is increased and the change in the hole density is
measured. $V_{\rm bg}$ is then reduced to recover the original
density. This procedure changes $E_{\perp}$ while maintaining the
same density to within 3\%, and allows calculation of the change in
$E_{\perp}$ from the way the gates affect the density.

In Fig.\ \ref{pic:w190:surf}(b) we show the Fourier spectra for the
measured magnetoresistance oscillations. Keeping in mind that we may
not expect a strict one-to-one correspondence between the
oscillatory density of states at the Fermi energy [Fig.\ 
\ref{pic:w190:surf}(a)] and the magnetoresistance oscillations
[Fig.\ \ref{pic:w190:surf}(b)] the agreement is very satisfactory.
However, these experimental and theoretical results indicate a
surprising discrepancy between $f_\pm^{\rm\,SdH}$ and $(h/e)N_\pm$.
In the following we will discuss possible explanations of these
results.

The common interpretation \cite{sto83} of SdH oscillations in the
presence of inversion asymmetry is based on the intuitive idea that
for small $B$ the Landau levels can be partitioned into two sets
which can be labeled by the two spin subbands. Each set gives rise
to an SdH frequency which is related to the population of the
respective spin subband according to Eq.\ (\ref{onsag}). However, a
comparison between the (partially) spin polarized eigenstates at
$B>0$ and the unpolarized eigenstates at $B=0$ shows that in general
such a partitioning of the Landau levels is not possible. This
reflects the fact that the orbital motion of up and down spinor
components is coupled in the presence of spin-orbit interaction,
i.e., it cannot be analyzed seperately.

\begin{figure}
\centerline{\epsfxsize=0.40\columnwidth\leavevmode
 \epsffile{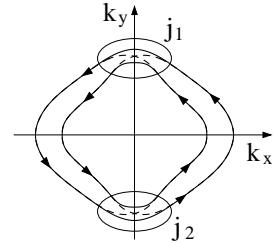}}\vspace{3mm}
\caption[]{\label{pic:w190:cont} Qualitative sketch of the
spin-split Fermi contours in ${\bf k}_\|$ space for a QW with growth
direction [113] (solid lines). In a simple semiclassical picture the
observation of a single peak near $E_\perp = 0$ in the Fourier
spectra of Fig.\ \ref{pic:w190:surf} can be explained by
trajectories in ${\bf k}_\|$ space which follow the dashed lines at
the junctions $j_1$ and $j_2$.}
\end{figure}

For many years, anomalous magneto-oscillations have been explained
by means of magnetic breakdown \cite{sho84}. In a sufficiently
strong magnetic field $B$ electrons can tunnel from an orbit on one
part of the Fermi surface to an orbit on another, separated from the
first by a small energy gap. The tunneling probability was found to
be proportional to $\exp(-B_0/B)$, with a breakdown field $B_0$,
similar to Zener tunneling \cite{sho84}. This brings into existence
new orbits which, when quantized, correspond to {\em additional}
peaks in the Fourier spectrum of the SdH oscillations. However, if
the anomaly of the SdH oscillations reported in Fig.\
\ref{pic:w190:surf} were due to magnetic breakdown, for $E_\perp =
0$ we would expect several frequencies $f^{\rm\, SdH}$ with
different values rather than the observed {\em single} frequency. In
a simple, semiclassical picture a single frequency could be
explained by two {\em equivalent} orbits in ${\bf k}_\|$ space as
sketched in Fig.\ \ref{pic:w190:cont}. However, the latter would
imply that the tunneling probabilities at the junctions $j_1$ and
$j_2$ are equal to one (and thus independent of $B$). We remark that
de Andrada e Silva {\em et al.}\ \cite{and94} studied anomalous
magneto-oscillations for spin-split electrons in a 2D system. Their
semiclassical analysis based on magnetic breakdown failed to predict
$B_0$ by up to a factor of three and $\Delta N$ by up to 17\%\ (see
Table III in Ref.\ \cite{and94}).

In order to understand the deviation from Eq.\ (\ref{onsag}) visible
in Fig.\ \ref{pic:w190:surf} we need to look more closely at
Onsager's semiclassical argument \cite{ons52} which is underlying
Eq.\ (\ref{onsag}). It is based on Bohr-Sommerfeld quantization of
the semiclassical motion of Bloch electrons, which is valid for
large quantum numbers. However, spin is an inherently quantum
mechanical effect, for which the semiclassical regime of large
quantum numbers is not meaningful. Therefore Bohr-Sommerfeld
quantization cannot be carried through in the usual way for systems
with spin-orbit interaction. In a semiclassical analysis of such
systems we have to keep spin as a discrete degree of freedom so that
the motion in phase space becomes a multicomponent vector field
\cite{lit92,bol98}, i.e., the motion along the spin-split branches
of the energy surface is coupled with each other and cannot be
analyzed seperately. In this problem geometric phases like Berry's
phase \cite{ber84} enter in an important way which makes the
semiclassical analysis of the motion of a particle with spin much
more intricate than the conventional Bohr-Sommerfeld quantization.

One may ask whether we can combine the older idea of magnetic
breakdown with the more recent ideas on Bohr-Sommerfeld quantization
in the presence of spin-orbit interaction. Within the semiclassical
theory of Ref.\ \cite{lit92} spin-flip transitions may occur at the
so-called mode-conversion points which are points of spin degeneracy
in phase space. Clearly these points are related to magnetic
breakdown. However, mode-conversion points introduce additional
complications in the theory of Ref.\ \cite{lit92} so that this
theory is not applicable in the vicinity of such points.

Clearly we can circumvent the complications of the semiclassical
theory by doing fully quantum mechanical calculations as outlined
above. We have performed extensive calculations and further
experiments which confirm that the results reported here are quite
common for 2D systems. In Ref.\ \cite{lu98} spin splitting of holes
was analyzed for two GaAs QW's which had only a front gate. Here
$V_{\rm fg}$ changes both the total density $N = N_{+} + N_{-}$ in
the well as well as the asymmetry of the confining potential. For
these QW's we obtain excellent agreement between the measured and
calculated frequencies $f_\pm^{\rm\,SdH}$ versus $N$ including the
observation of a single SdH frequency near $N=3.8\cdot
10^{11}$~cm$^{-2}$ when the QW becomes symmetric. However, there is
again a significant discrepancy between $\Delta f$ and $(h/e) \Delta
N$.

Our results apply to other III-V and II-VI semiconductors whose band
structures are similar to GaAs in the vicinity of the fundamental
gap \cite{lan82}. Our calculations indicate that the deviations from
Eq.\ (\ref{onsag}) are related to the anisotropic terms in the
Hamiltonian. If the Hamiltonian is axially symmetric Eq.\
(\ref{onsag}) is fulfilled. This is consistent with the
semiclassical analysis of spin-orbit interaction in Ref.\
\cite{lit92} where it was found that in three dimensions no Berry's
phase occurs for spherically symmetric problems. We note that for
holes in 2D systems the anisotropy of ${\cal E}_\pm ({\bf k}_\|)$ is
always very pronounced \cite{eke85}. It is also a well-known feature
of the Hamiltonian for electrons, in particular for semiconductors
with a larger gap \cite{ros89}. Up to now most experiments have
analyzed spin splitting and SdH oscillations for 2D electron systems
\cite{wol89,luo90,nit97}. To lowest order in ${\bf k}$ the SIA
induced spin splitting in these systems is given by the so-called
Rashba term \cite{bych84,ros89} which has axial symmetry. For this
particular case it can be shown analytically that Eq.\ (\ref{onsag})
is fulfilled.

\begin{figure}
\centerline{{\epsfxsize=0.68\columnwidth\leavevmode
 \epsffile{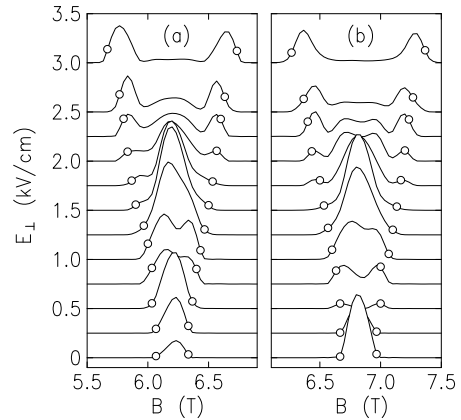}}}\vspace{3mm}
\caption[]{\label{pic:w150:110} Calculated Fourier
spectra of the SdH oscillations versus magnetic field $B$ for different
values of electric field $E_\perp$ for a 150 {\AA} wide
GaAs-Al$_{0.5}$Ga$_{0.5}$As QW with crystallographic growth
direction [110] and 2D hole densities (a) $N = 3.0 \cdot
10^{11}$~cm$^{-2}$ and (b) $N = 3.3 \cdot 10^{11}$~cm$^{-2}$. The
open circles show the expected Fourier transform peak positions
$(h/e)N_\pm$ according to the calculated spin splitting $N_\pm$ at
$B=0$.}
\end{figure}

For different crystallographic growth directions spin splitting and
SdH oscillations behave rather differently. Moreover, these
quantities depend sensitively on the total 2D hole density $N=N_+ +
N_-$ in the well. In Fig.~\ref{pic:w150:110} we have plotted the
calculated SdH Fourier spectra versus $E_\perp$ for a GaAs QW with
growth direction [110] and $N=3.0 \cdot 10^{11}$ cm$^{-2}$ [Fig.\
\ref{pic:w150:110}(a)] and $N=3.3 \cdot 10^{11}$ cm$^{-2}$ [Fig.\
\ref{pic:w150:110}(b)]. Open circles mark the expected peak
positions $(h/e)N_\pm$ according to the spin splitting $N_\pm$ at
$B=0$ \cite{totdens,asym}.  Again, the peak positions in the Fourier
spectra differ considerably from the expected positions
$(h/e)N_\pm$. Close to $E_\perp =0$ there is only one peak at
$(h/2e)N$. Around $E_\perp = 1.0$ kV/cm we have two peaks, but at
even larger fields $E_\perp$ the central peak at $(h/2e)N$ shows up
again. At $E_\perp \approx 2.25$ kV/cm we have a triple peak
structure consisting of a broad central peak at $(h/2e)N$ and two
side peaks at approximately $(h/e)N_\pm$. In Fig.~\ref{pic:w150:110}
we have a significantly smaller linewidth than in
Fig.~\ref{pic:w190:surf}. Basically, this is due to the fact that
for the Fourier transforms shown in Fig.\ \ref{pic:w150:110} we used
a significantly larger interval of $B^{-1}$ (10.0 T$^{-1}$ as
compared with 3.83 T$^{-1}$) in order to resolve the much smaller
splitting for growth direction [110]. We note that for $E_\perp = 0$
the SdH oscillations are perfectly regular over this large range of
$B^{-1}$ with just {\em one} frequency, which makes it rather
unlikely that the discrepancies between $\Delta f$ and $(h/e)\Delta
N$ could be caused by a $B$ dependent rearrangement of holes between
the Landau levels.

Similar results like those shown in Figs.\ \ref{pic:w190:surf} and
\ref{pic:w150:110} have been obtained also for growth direction
[001], but the spectra were more complicated with, e.g., several SdH
frequencies for $E_\perp = 0$. Our calculations for holes are based
on the fairly complex multiband Hamiltonian of Ref.\ \cite{trr79}.
We obtained qualitatively the same results by analyzing the simpler
$2\times 2$ Hamiltonian of Ref.\ \cite{ros89}. However, this model
is appropriate for electrons in large-gap semiconductors, where spin
splitting is rather small, so that it is more difficult to observe
these effects experimentally.

In summary, we have both measured and calculated the SdH
oscillations of 2D hole systems in GaAs QW's. As opposed to the
predictions of a semiclassical argument due to Onsager, we conclude
that the $B = 0$ spin splitting is not simply related to the SdH
oscillations at low magnetic fields \cite{prop}. This is explained
by the inapplicability of conventional Bohr-Sommerfeld quantization
for systems with spin-orbit interaction.

R.\ Winkler benefitted from stimulating discussions with O.\
Pankratov, M.\ Suhrke and U.\ R\"ossler, and he is grateful to S.\
Chou for making available his computing facilities. The work at
Princeton University is supported by the NSF and ARO. M.\ Shayegan
acknowledges support by the Alexander von Humboldt Foundation.

\end{multicols}

\begin{thebibliography}{99}
\bibitem{totdens} In both our calculations and experiments we always
 find that the sum of the two SdH frequencies correctly give the
 total density, i.e., $N = (e/h) (f_+^{\rm\,SdH} + f_-^{\rm\,SdH})$.
 Throughout the paper we consider the case that only one pair of
 spin-split subbands is occupied.
\bibitem{asym} The asymmetry of the Fourier spectra and the spin
 splitting visible in Fig.\ \ref{pic:w190:surf} for $E_\perp <0$
 compared with $E_\perp > 0$ is due to the low-symmetry growth
 direction [113] of our sample. For growth direction [110] (Fig.\
 \ref{pic:w150:110}) $E_\perp <0$ and $E_\perp >0$ are equivalent.
\bibitem{beat} The calculated values of $\Delta N$ near $E_\perp = 0$
 would imply a beating pattern of the SdH oscillations with two
 nodes within the investigated range of $B$.
\bibitem{prop} We wish to propose that the $B=0$ spin splitting of 2D
 holes may be measured using inelastic spin-flip Raman scattering
 similar to the experiments for electrons in Ref.\ \cite{jus92}.
 Such experiments are planned.
\end{thebibliography}
\end{document}